\documentclass[prb,eqsecnum,preprint,showpacs]{revtex4}

\newcounter{xitem}
\usepackage{bm}
\usepackage{stmaryrd}
\usepackage{amsmath}
\usepackage{amssymb}
\usepackage{graphicx}
\usepackage{textcomp}
\usepackage{calrsfs}
\usepackage{yfonts}
\newcommand{\be}{\begin{equation}}
\newcommand{\ee}{\end{equation}}
\newcommand{\ben}{\begin{equation*}}
\newcommand{\een}{\end{equation*}}
\newcommand{\bea}{\begin{eqnarray}}
\newcommand{\eea}{\end{eqnarray}}

\newcommand{\yitem}{\refstepcounter{xitem}\item[{\bf\arabic{xitem}.}]}
\begin{document}

\title{Resource Letter VWCPF-1: Van der Waals and Casimir-Polder forces}


\author{Kimball A. Milton}
\affiliation{Homer L. Dodge Department of
Physics and Astronomy, University of Oklahoma, Norman, OK 73019-2061, USA;
e-mail: milton@nhn.ou.edu}

\begin{abstract}
This Resource Letter provides a guide to the literature on van der Waals and
Casimir-Polder forces.  Journal articles, books, and other documents are cited
on the following topics: nonretarded or van der Waals forces, retarded dispersion forces or
Casimir-Polder forces between atoms or molecules, Casimir-Polder forces between a molecule
and a dielectric or conducting body, the summation of Casimir-Polder forces as
leading to the Casimir and Lifshitz forces between conducting and dielectric
bodies, Casimir friction, applications to nanotechnology, 
the nature of the quantum vacuum, and experimental tests of the theory
of Casimir and Casimir-Polder and van der Waals forces.
\end{abstract}

\pacs{42.50.Lc, 32.10.Dk, 12.20.-m, 11.80.La}
\maketitle
\section{Introduction}
The forces between molecules, first inferred from deviations from the ideal
gas laws, are often due to quantum fluctuations.  These can be thought of
as fluctuations of the molecular or atomic
electric-dipole moments themselves, either permanent or induced, or as 
fluctuations in the electromagnetic fields surrounding
the dipoles.  These forces were calculated from first principles starting in
the 1930s in the nonretarded regime (that is, the finiteness of the 
speed of light being neglected) where the forces are called van der Waals, 
and in 1947 in the retarded regime (for distances large compared to
the speed of light times a characteristic time) where they are referred to
as Casimir-Polder (CP) forces.  Progress in the subject has
continued at an accelerating pace since then.  The applications are important
in atomic and molecular physics, quantum field theory, and nanotechnology.
For example, Casimir attraction can give rise to ``stiction'' 
(frictional forces that cause parts to stick together) between the
micro-components, which will cause the device to fail, or, more excitingly,
Casimir forces can drive the operation of nanomachinery.
Van der Waals and Casimir-Polder forces may be thought of as special cases
of, or as the origin of
the general Casimir or quantum vacuum forces between macroscopic and mesoscopic
bodies, and provide an important window into the quantum fluctuating nature of reality.

It should be noted that the author has been highly selective in choosing representative
articles in this rapidly developing field.  Because the central focus is on van
der Waals and Casimir-Polder forces involving atoms, many topics of great importance
in the general field of quantum vacuum energy phenomena, such as the dynamical
Casimir effect, the coupling of Casimir energy to gravity or to modified gravity,
and generally, Casimir energies in curved spacetime, have been excluded.

In this Resource Letter we use Gaussian units.  To convert to SI units,
replace the Gaussian polarizability $\alpha$
by $\alpha^{\rm SI}/4\pi\epsilon_0$.
and replace the Gaussian permittivity $\varepsilon$
by the relative permittivity $\varepsilon^{\rm SI}/\epsilon_0$.

\section{Books, Popular Articles, and Reviews}
Quantum vacuum forces are discussed in the following books and monographs.
\begin{itemize}
\yitem  \label{milonni}
{\bf The Quantum Vacuum: An Introduction to Quantum Electrodynamics},
  P.~W.~Milonni (Academic Press, Boston, 1994). (I)
\yitem {\bf The Casimir Effect and Its Applications}, V. M. Mostepanenko
and N. N. Trunov (Oxford University Press, Oxford, 1997). (A)
\yitem {\bf The Casimir Effect: Physical Manifestations of Zero-Point
Energy}, K. A. Milton (World Scientific, Singapore, 2001). (A)
\yitem  {\bf Advances in the Casimir Effect},
 M.~Bordag, G.~L.~Klimchitskaya, U.~Mohideen, and V. M. Mostepanenko
(Oxford University Press, Oxford, 2009). (A)
\end{itemize}
Somewhat more specialized are books with a molecular orientation:
\begin{itemize}
 \yitem {\bf Dispersion Forces}, J. Mahanty and B. W. Ninham
(Academic Press, London, 1976). (I)
\yitem {\bf Molecular Quantum Electrodynamics: An Introduction
to Radiation-Molecule Interactions}, D. P. Craig and T.
Thirunamachandran (Academic Press, London. 1984). (I)
\yitem {\bf Intermolecular and Surface Forces}, J. Israelachvili
(Academic Press, New York, 1985). (I)
\yitem {\bf Molecular Quantum Electrodynamics: Long-range
Intermolecular Interactions}, A. Salam (Wiley, Hoboken, NJ, 
2010). (I)
\end{itemize}
Very useful books of a more mathematical character are
the following:
\begin{itemize}
\yitem {\bf Zeta Regulation Techniques with Applications},
E. Elizalde, S. D. Odintsov, and A. Romeo (World Scientific,
Singapore, 1994). (A)
\yitem {\bf Ten Physical Applications of Spectral Zeta Functions}
(Lecture Notes in Physics),
E. Elizalde (Springer, Berlin, 1995). (A)
\yitem {\bf Spectral Functions in Mathematics and Physics},
K. Kirsten  (Chapman \& Hall/CRC Press, Boca Raton, FL, 2002). 
\end{itemize}
The following are a selection of review articles:
\begin{itemize}
\yitem ``The Casimir Effect,'' G. Plunien, B. M\"uller, and W. Greiner,
Phys.\ Rep.\ {\bf 134}, 87--193 (1986). (I)
\yitem  ``The Casimir Effect and Its Applications,''
  V.~M.~Mostepanenko and N.~N.~Trunov,
   Sov.\ Phys.\ Usp.\  {\bf 31}, 965--987 (1988). (I)
\yitem ``New Developments in the Casimir Effect,''
M.~Bordag, U.~Mohideen, and V.~M.~Mostepanenko,
    Phys.\ Rept.\  {\bf 353}, 1--205 (2001). (A)
\yitem   \label{miltonrev}
``The Casimir Effect: Recent Controversies and Progress,''
K.~A.~Milton,  J.\ Phys.\ A {\bf 37}, R209--R277 (2004). (A)
\yitem ``Dispersion Forces in Macroscopic Quantum Electrodynamics.''
S. Y. Buhmann and D.-G. Welsch,
Prog.\ Quant.\ Electron.\ {\bf31}, 51--130 (2007). (I) 
\yitem   ``The Casimir Force Between Real Materials: Experiment and Theory,''
G.~L.~Klimchitskaya, U.~Mohideen, and V.~M.~Mostepanenko,
   Rev.\ Mod.\ Phys.\  {\bf 81}, 1827--1885 (2009). (A)
\end{itemize}
A Resource Letter on the Casimir Effect was published over a decade ago.
\begin{itemize}
  \yitem 
``Resource Letter CF-1: Casimir Force,''
S. K. Lamoreaux,
Am.\ J. Phys.\ {\bf 67}, 850--861 (1999). (E)
\end{itemize}

The following give historical perspective:
\begin{itemize}
\yitem {\bf Physics in the Making: Essays on Developments in 20th
Century Physics in Honour of H. B. G. Casimir on the Occasion of his
80th Birthday}, A. Sarlemijn and M. J. Spaarnay (North Holland,
Amsterdam, 1989). (E)
\yitem ``Some Remarks on the History of the So-Called Casimir Effect,''
 H. B. G. Casimir, in {\bf Casimir Effect 50 Years Later}, ed.\
 M. Bordag (World Scientific, Singapore 1999), pp.~3--9 [Proceedings of the
Fourth Workshop on Quantum Field Theory Under the Influence of External
Conditions, Leipzig, 1998]. (E)
\yitem ``Casimir Effects in Atomic, Molecular, and Optical Physics,''
J. F. Babb, Advances in Atomic, Molecular, and Optical Physics {\bf 59},
1--20 (2010). (E)
\end{itemize} 
The QFEXT series of conferences give a continuing overview.  The latest
proceedings is the following.
\begin{itemize}
\yitem {\bf Proceedings of the Ninth Conference on Quantum Field Theory
Under the Influence of External Conditions (QFEXT09)}, ed.~K. A. Milton
and M. Bordag (World Scientific, Singapore, 2010). (A)
\end{itemize}
Other recent collections of papers include the following:
\begin{itemize}
\yitem {\bf Cosmology, the Quantum Vacuum, and Zeta Functions: A
Choice of Papers}, eds.~E. Elizalde and S. D. Odintsov (Tomsk State
Pedagogical University, Tomsk, Russia, 2009). (A)
\yitem {\bf Cosmology, Quantum Vacuum, and Zeta Functions}, eds.~D.
S\'aez-G\'omez, S. D. Odintsov, and S. Xamb\'o (Springer, Berlin, 2011). (A)
\end{itemize}
Popular articles include:
\begin{itemize}
\yitem ``The Force Between Molecules,'' B. V. Derjaguin, Sci.\ Am.\
{\bf203} (7), 47--53 (1960). (E)
\yitem ``Retarded, or Casimir, Long-Range Potentials,'' L. Spruch,
Physics Today {\bf 39} (11) 37--45 (1986). (E)
\yitem ``Essentials of the Casimir Effect and its Computation,''
E. Elizalde and A. Romeo, Am.\ J. Phys.\ {\bf 59}, 711--719 (1991). (I)
\yitem ``Long Range (Casimir) Interactions,'' L. Spruch, Science {\bf 272},
1452--1455 (1996). (E)
\yitem``A Maritime Analogy of the Casimir Effect,'' S. L. Boersma, Am.\ J.
Phys.\ {\bf 64}, 539--541 (1996). (E)
\yitem ``Experiment and Theory in the Casimir Effect,'' G. L. Klimchitskaya
and V. M. Mostepanenko, Contemp.\ Phys.\ {\bf 47}, 131--144 (2006). (E)
\yitem  ``Casimir Forces: Still Surprising After 60 Years,'' 
S. K. Lamoreaux, Physics Today {\bf 60} (2), 40--45 (2007). (E)
\yitem ``Casimir Effects,'' P. W. Milonni, Phys.\ Scr.\ {\bf 76}, C167--C171
(2007). (E)
\yitem ``Casimir Force Could Drive Tiny Ratchets,'' 
H. Johnston, Physics World, May 2, 2007. (E)
\yitem ``Casimir Effect Goes Negative,'' E. Cartlidge, Physics World, 
Jan.~8, 2009. (E)
\yitem ``Casimir Forces between Solids Can Be Repulsive,''
J. Miller, Physics Today {\bf 62} (2), 19--21. (E)
\yitem ``Thermal Casimir Force Seen for the First Time,''
H. Johnston, Physics World, Feb.~8, 2011. (E)

\end{itemize}
  
\section{van der Waals to Casimir-Polder}
Van der Waals forces between molecules were first inferred from deviations from
the ideal gas laws.
\begin{itemize}
\yitem {\bf O ver de Continue\"iteit van den Gas- en Vloeistoftocstand},
J. D. van der Waals (Ph.D. thesis, Leiden, 1873). (E)
\yitem {``O ver de Continuiteit van den Gas- en Vloeistofiocstand 
Academisch Proefschrift'' (book review)}, J. C. Maxwell, 
Nature {\bf 10}, 477--480 (1874). (I)
\end{itemize}
But quantum mechanics was required to understand how such forces could arise
from neutral atoms or molecules.  In the short-distance, or nonretarded regime,
where the speed of light is regarded as infinite, 
London derived, using second-order
perturbation theory, the interaction between two fluctuating dipoles,
characterized by a polarizability $\alpha$. That is, the dipole moment $\bf p$ 
is proportional to the electric field $\bf E$, $\mathbf{p}(\omega)=
\alpha(\omega)\mathbf{E}(\omega)$.
The resulting interaction energy fell
off with the distance $r$ between the atoms like $r^{-6}$,
\begin{equation}
V_{\rm vdW}=-\frac3\pi\frac\hbar{r^6}\int_0^\infty d\zeta\alpha_1(i\zeta)
\alpha_2(i\zeta),\label{london}
\end{equation}
where $\alpha_i$ is the polarizability of the $i$th atom, as a function of
the imaginary frequency $\omega=i\zeta$  (that is, we rotate $\omega$ to
the imaginary axis, so $\zeta$ is real).
\begin{itemize}
 \yitem {``Zur Theorie und Systematik der Molekularkr\"afte,''} 
F. London, Z. Physik {\bf 63}, 245--279 (1930). (I)
\yitem {``\"Uber das Verh\"altnis der Waalsschen Kr\"afte zu
den hom\"oopolarn Bindungskr\"aften,''}
R. Eisenschitz and F. London, Z. Phys.\ {\bf 60}, 491--527 (1930). (I)
\end{itemize}
This, however, only holds for distances of order 10 nm or less.  For longer
distances, the finite speed of light must be taken into account, that is,
retardation must be considered.  Casimir and Polder 
showed that the interaction energy between two isotropic
polarizable  atoms falls off faster, like $r^{-7}$,
at zero temperature, in terms of the static, zero-frequency, polarizabilities
$\alpha_i=\alpha_i(0)$,
\begin{subequations}
\be
V_{\rm CP}=-\frac{23}{4\pi}\frac{\alpha_1\alpha_2}{r^6}\frac{\hbar c}{r},
\label{cp}
\ee 
whereas, as was later realized, 
for high temperature, the dependence on $\hbar c/r$ is replaced
by $kT$, where $k$ is Boltzmann's constant,
\be
V_{\rm CP}\sim-\frac{\alpha_1\alpha_2}{r^6} kT,\quad T\to\infty.\label{cpt}
\ee
\end{subequations}
Casimir and Polder 
reproduced the London result at short distances, where the transition
between the retarded and nonretarded regimes is set by some characteristic
wavelengths of the relevant atomic transitions.
\begin{itemize}
 \yitem \label{cppaper}
{``The Influence of Retardation on the London-van der Waals Forces,''}
H. B. G. Casimir and D. Polder, Phys. Rev. {\bf 73}, 360--372 (1948). (I)
\end{itemize}

In this paper, the authors also derived the interaction
energy between an atom, of static polarizability $\alpha$, 
and a metallic plate a distance $Z$ away,
\be
E_{\rm CP}=-\frac{3\alpha}{8\pi Z^4}\hbar c.\label{cp0}
\ee

As a result of a conversation with Bohr, Casimir realized that these results
could be obtained more simply by considering the zero-point fluctuations of
the electromagnetic fields between the atoms, or the atom and the plate, as he
explained in a lecture in Paris:
\begin{itemize}
 \yitem ``Sur les Forces van der Waals-London,'' H. B. G. Casimir, in 
{\bf Colloque sur la th\'eorie de la liaison chimique},
Paris, 12--17 April 1948, published in J. Chim.\ Phys.\ 
et Phys.\ Chim.\ Biol.\ {\bf 46}, 407--410 (1949). (I)
\end{itemize}

If the atoms are not isotropic (spherically symmetric) the interaction depends
on the orientation between the polarization tensors of the atoms.  Thus, if
we assume only linearity, so that there is a tensor (dyadic) polarization,
\begin{equation}
\mathbf{p=\bm{\alpha}\cdot E},
\end{equation}
the interaction between such an atom and an isotropic slab a distance $Z$ away
is generalized to
\begin{equation}
E_{\rm CP}=-\frac{\mbox{Tr\,}\bm{\alpha}}{8\pi Z^4}.
\end{equation}
And the Casimir-Polder potential between two anisotropic atoms is generalized
to
\be
V_{\rm CP}=-\frac1{4\pi r^7}\left[\frac{13}2\mbox{Tr\,}\bm{\alpha}_1\cdot
\bm{\alpha}_2-28\,\mbox{Tr\,}(\bm{\alpha}_1\cdot\mathbf{\hat{r}})\, 
(\bm{\alpha}_2\cdot\mathbf{\hat{r}})
+\frac{63}2(\mathbf{\hat r}\cdot\bm{\alpha}_1\cdot\mathbf{\hat r})\,
(\mathbf{\hat r}\cdot\bm{\alpha}_2\cdot\mathbf{\hat r})\right],
\ee
where $\mathbf{r}=r\mathbf{\hat r}$ is the relative position vector of the
two atoms, and $\mbox{Tr}$ stands for the trace of the matrix.  
This generalizes the form given in Ref.~[\ref{cppaper}] when
the two polarizations are not simultaneously diagonalizable.

For a pedagogical non-field-theoretical rederivation of van der Waals
and Casimir-Polder forces, see
\begin{itemize}
\yitem ``The van der Waals Interaction,''
B. R. Holstein, Am.\ J. Phys.\ {\bf69}, 441--449 (2001). (I)
\end{itemize}  See also Sec.~\ref{sec:devcp}.
For interactions between a polarizable atom and a multilayer substrate, see
\begin{itemize}
\yitem ``van der Waals and Retardation (Casimir) Interactions of an Electron 
or an Atom with Multilayered Walls,'' F. Zhou and L. Spruch,
Phys.\ Rev.\ A {\bf52}, 297--310 (1995). (I)
\end{itemize} 
 
\section{Casimir effect}\label{sec:ce}
The realization by Casimir that Casimir-Polder forces could be
understood in terms of quantum field fluctuations
 allowed him to immediately recognize that there would be
a quantum-electrodynamic interaction between two neutral  parallel 
conducting plates, which resulted in an attractive force 
$F$ between those plates 
having area $A$ and separation distance $a$:
\be
\frac{F}{A}=-\frac{\pi^2\hbar c}{240 a^4}.
\ee
\begin{itemize}
 \yitem {``On the Attraction Between Two Perfectly
Conducting Plates,''} H. B. G. Casimir, Proc.\ Kon.\ Ned.\ Akad.\ Wetensch.\ 
{\bf 51}, 793--795 (1948). (I)
\end{itemize}
Numerically, this force is quite small,
\be
\frac{F}A=-1.30\times 10^{-27} \mbox{ N \,m}^2a^{-4},
\ee
but quite significant at the 100 nm scale.

A few years later, Lifshitz generalized the Casimir force between perfectly
conducting plates to the force between dielectric slabs.
\begin{itemize}
\yitem {``The Theory of Molecular Attractive Forces
Between Solids,''} E. M. Lifshitz, Zh.\ Eksp.\ Teor.\ Fiz., {\bf 29}, 94--110
(1955) [English transl.: Soviet Phys.\ JETP {\bf 2}, 73--83 (1956)]. (I)
\yitem {``General Theory of van der Waals' Forces,''} 
I. D. Dzyaloshinskii, E. M. Lifshitz, and L. P.
Pitaevskii, Usp.\ Fiz.\ Nauk {\bf 73}, 381--422 (1961) [English transl.: Soviet
Phys.\ Usp. {\bf 4}, 153--176 (1961)]. (A)
\yitem {\bf Electrodynamics of Continuous Media}, 
L. D. Landau and E. M. Lifshitz (Pergamon, Oxford, 1960). (I)
\end{itemize}
The Lifshitz formula may be easily understood in terms of multiple
reflections between the parallel surfaces; an accessible derivation is
found in Ref.~[\ref{miltonrev}].  For the case of two thick parallel slabs,
with dielectric constants $\varepsilon_1$, $\varepsilon_2$, respectively,
separated by a medium having dielectric constant $\varepsilon_3$ and
thickness $a$, the force/area at zero temperature is given in terms
of integrals over imaginary frequency $\zeta$ and wave number $k$ (both
with dimensions of inverse length):
\begin{equation}
\frac{F}A=-\frac{\hbar c}{8\pi^2}\int_0^\infty 
d\zeta \int_0^\infty dk^2 2\kappa_3\left(
d_{\rm TE}^{-1}+d_{\rm TM}^{-1}\right),
\end{equation}
where for the ``transverse electric'' polarization
\begin{equation}
d_{\rm TE}=\left(r^{\rm TE}_{13}\right)^{-1}\left(r^{\rm TE}_{23}\right)^{-1} 
e^{2\kappa_3a}-1,
\end{equation}
in terms of the reflection coefficients for a single dielectric interface,
\begin{equation}
r^{\rm TE}_{ij}=\frac{\kappa_i-\kappa_j}{\kappa_i+\kappa_j}.
\end{equation}
Here $\kappa_i^2=k^2+\varepsilon_i\zeta^2$.  For the TM mode, one merely 
replaces $\kappa_i\to\kappa_i/\varepsilon_i$ in the reflection coefficients.

In these papers, Lifshitz and collaborators showed that when the
dielectric constants of the media became large, the result of Casimir for
perfect conductors was recovered. They further showed
 that when the media were tenuous, so when 
\be
\varepsilon-1=4\pi N\alpha\ll1,
\ee
where $N$ is the number density of polarizable molecules in the media,
the Lifshitz force could be understood as the pairwise summation of 
Casimir-Polder forces between the molecules, and thereby rederived 
Eq.~(\ref{cp}).  Many years later, these results and more were rederived
by Schwinger and his postdocs.
\begin{itemize}
\yitem \label{sdm}{``Casimir Effect in Dielectrics,''} 
J. Schwinger, L. L. DeRaad, Jr., and K. A. Milton, Ann.\ Phys.\ (N.Y.) 
{\bf115}, 1--23 (1978). (A)
\end{itemize}

In this paper, as in the earlier Lifshitz papers, temperature dependence
was also discussed.  However, in order to recover the expected result for
perfectly conducting plates, a prescription was adopted asserting that formally
the limit $\varepsilon\to\infty$ had to be imposed before taking the zero
frequency limit, which arises in the Matsubara sum over discrete frequencies
that occurs for finite temperature $T$.  That is, to pass from the 
zero-temperature Lifshitz formula to the finite-temperature expression,
one replaces the frequency integral by a discrete sum:
\begin{subequations}
\be
\int_{-\infty}^\infty \frac{d\zeta}{2\pi}f(\zeta)\to kT\sum_{m=-\infty}^\infty
f(\zeta_m),
\ee where the Matsubara frequency is
\begin{equation}
\zeta_m=2\pi m k T.
\end{equation}  
\end{subequations}
(Here $m$ is a positive or negative
integer, including 0.) This prescription
was designed to achieve agreement with the ideal metal
limit, and 
guaranteed that the result was consistent with the
third law of thermodynamics, that the entropy should vanish at zero 
temperature, although this was not widely appreciated at the time, and
was not explicitly stated in the papers.
But this prescription came under serious scrutiny at the
dawn of the 21st
century, and is now generally regarded as erroneous---see Sec.~\ref{sec:con}.

It should also be noted that there have been various attempts to divorce
Casimir forces from the concept of zero-point energy.  Notable among these
are the following.
\begin{itemize}
\yitem ``Casimir Effect in Source Theory,'' J. Schwinger, Lett.\ Math.\
Phys.\ {\bf 1}, 43--47 (1975). (I)
\yitem ``The Casimir Effect and the Quantum Vacuum,''
R.L. Jaffe, Phys.\ Rev.\ D {\bf72},  021301 (2005). (I)
\end{itemize} See also Refs.~[\ref{sdm}] and [\ref{milonni}].
However, it is the view of the author that since the predictions of 
the fluctuating field description built into quantum field theory
are unaltered by these reformulations, there is no physical significance
to these distinctions.  The physical arena where these distinctions
might be relevant is in connection with dark energy---See Sec.~\ref{concl}.

There were many important papers on the Casimir effect published in the 
thirty-year period 1960--1990.  Among these, special note should be given to
the following.
\begin{itemize}
 \yitem ``Vacuum Stress between Conducting Plates: An Image Solution,''
L. S. Brown and G. J. Maclay, Phys.\ Rev.\ {\bf184}, 1272--1279 (1969). (I)
\end{itemize}
Brown and Maclay extracted the vacuum expectation value of the stress
tensor that expresses both the stress (forces) on the plates as well as
the local energy density,
\begin{equation}
\langle T^{\mu\nu}\rangle= u\left(4\hat z^\mu\hat z^\nu-g^{\mu\nu}\right),
\end{equation}
where $u$ is the energy density between the plates, separated by a distance $a$,
\begin{equation}
u=-\frac{\pi^2\hbar c}{720 a^4}.
\end{equation}
The stress tensor is expressed in terms of the metric tensor of flat space, 
$g^{\mu\nu}=\mbox{diag}(-1,1,1,1)$, and where $\hat z^\mu$ is the unit
vector in the $z$ direction.  Note that this says that the quantum
energy density is constant between the plates (and zero outside).
They further performed a full derivation of the Casimir energy at
finite temperature between two parallel plates in the framework
of thermal field theory, and observed a symmetry between low and high
temperature behaviors, a type of duality.

\section{Experiment}
Experimental confirmation of the Casimir force took a long time in coming.
Early experiments in the 1950s and 1960s were relatively inconclusive,
the best of which did ``not contradict Casimir's theoretical prediction.''
\begin{itemize}
\yitem ``Direct Measurement of the Molecular Attraction of Solid
Bodies. I. Statement of the Problem and Measurement of the Force
by Using Negative Feedback,''
B. V. Deriagin (Derjaguin) and I. I. Abrikosova,
 Zh.\ Eksp.\ Teor. Fiz.\ {\bf 30} 993 (1956)
[English transl.:  Soviet Phys. JETP {\bf 3}, 819--829 (1957)]. (I)
\yitem ``Measurements of Attractive Forces between Flat Plates,'' 
M. Sparnaay, Physica {\bf24}, 751--764 (1958). (I) 
\end{itemize}
The Lifshitz theory was convincingly confirmed by Sabisky and Anderson.
\begin{itemize}
\yitem {``Verification of the Lifshitz Theory of the van der Waals Potential 
Using Liquid-Helium Films,''} E. S. Sabisky and C. H. Anderson, Phys.\
Rev.\ A {\bf 7}, 790--806 (1973). (I)
\end{itemize}
They attributed the force of adhesion of a helium film to a substrate in a
helium atmosphere in terms of the Lifshitz theory.  The first convincing
measurement of the Casimir force between conducting plates did not appear
until 1997:
\begin{itemize}
\yitem {``Demonstration of the Casimir Force in the 0.6 to 6 $\mu$m Range,''} 
S. K. Lamoreaux, Phys.\ Rev.\ Lett.\ {\bf 78}, 5--8 (1997). (I)
\end{itemize}
This, and almost all subsequent Casimir measurements, involved not two
parallel plates (which are exceedingly difficult to keep parallel), but
a plane and a spherical lens (or a sphere), which force, when the separation
distance between two objects is very small, could be deduced from that for
parallel plates by the ``proximity force approximation.''
\begin{itemize}
\yitem {``Untersuchungen \"uber die Reibung und Adh\"asion, IV. Theorie
des Anhaftens kleiner Teilchen,''} 
B. V. Derjaguin, Kolloid Z. {\bf 69}, 155--164 (1934).
(I) \label{ref:pfa}
\end{itemize}
In this approximation, the force between a spherical conductor of
radius $R$ and
a conducting plane separated by a distance $d$ is
\begin{equation}
F=-\frac{\pi^3}{360}\frac{R}d\frac{\hbar c}{d^2},\quad R\gg d.
\end{equation}
It has proved impossible to reliably extend this approximation beyond
the regime where the two objects are nearly touching.  Only in recent
years have exact and reliable numerical calculations become possible.
(See Sec.~\ref{secexact}.) Subsequent, more refined observations of the 
Casimir force include the following.
\begin{itemize}
\yitem {``Precision Measurement of the Casimir Force from 0.1 to 0.9
$\mu$m,''} U. Mohideen and A. Roy,
Phys. Rev. Lett. {\bf81}, 4549--4552 (1998). (A)
\yitem {``Measurement of the Casimir Force between Dissimilar Metals,''}
R. S. Decca, D. L\'opez, E. Fischbach, and D. E. Krause,
Phys.\ Rev.\ Lett.\ {\bf91}, 050402 (4 pages) (2003). (I)
\yitem \label{decca05}
``Precise Comparison of Theory and New Experiment for the Casimir 
Force Leads to Stronger Constraints on Thermal Quantum Effects and Long-Range 
Interactions,'' R. S. Decca, D. L\'opez, E. Fischbach, G. L. Klimchitskaya, 
D. E. Krause, and V. M. Mostepanenko, Ann.\ Phys.\ {\bf318},  37--80 (2005).
(A)
\yitem \label{decca07}
``Tests of New Physics from Precise Measurements of the Casimir Pressure 
Between Two Gold-Coated Plates,'' R. S. Decca, D. L\'opez, E. Fischbach, 
G. L. Klimchitskaya, D. E. Krause, and V. M. Mostepanenko,
 Phys.\ Rev.\ D {\bf75}, 077101 (4 pages) (2007). (A)
\yitem \label{decca07a}
``Novel Constraints on Light Elementary Particles and 
Extra-Dimensional Physics from the Casimir Effect,''
R.S. Decca, D. L\'opez, E. Fischbach, G.L. Klimchitskaya, D.E. Krause, and
V. M. Mostepanenko, Eur.\ Phys.\ J. C {\bf 51}, 963-975 (2007).  (A)
\yitem ``Measurement of the Casimir Force Between Parallel Metallic Surfaces,''
G. Bressi, G. Carugno, R. Onofrio, and G. Ruoso,
Phys.\ Rev.\ Lett.\ {\bf88}, 041804 (4 pages) (2002). (I)
\end{itemize}
The last reference is the only measurement of the Casimir force carried out
between parallel surfaces, and consequently, the systematic errors are 
relatively large owing to the difficulty of maintaining parallelism.
Electrostatic calibration is a problem in all these measurements,
since the Casimir forces may be overwhelmed by electrostatic forces.
\begin{itemize}
\yitem 
``Progress in Experimental Measurements of the Surface-Surface Casimir Force: 
Electrostatic Calibrations and Limitations to Accuracy,''
S. K. Lamoreaux,  arXiv:1008.3640, invited review paper to appear in a Lecture Notes in 
Physics Volume on Casimir physics edited by Diego Dalvit, Peter Milonni, 
David Roberts, and Felipe da Rosa. (I)
\end{itemize}

The Casimir-Polder force between an atom and a conducting plate until
recently had only been convincingly observed in a single experiment by 
Hinds' group:
\begin{itemize}
\yitem \label{sukenik} ``Measurement of the Casimir-Polder force,'' 
 C. I. Sukenik, M. G. Boshier, D. Cho, V. Sundoghar, 
and E. A. Hinds, Phys.\ Rev.\ Lett. {\bf 70}, 560--563 (1993). (I)
\end{itemize}
This experiment actually measured the force between an atom and
a pair of plates in a wedge configuration. This 
was compared to the force for an atom
between parallel plates, which had been worked out by Barton:
\begin{itemize}
 \yitem {``Quantum-Electrodynamic Level Shifts between Parallel Mirrors: 
Applications, Mainly to Rydberg States,''} G. Barton, 
Proc.\ Roy.\ Soc.\ London {\bf 410}, 175--200 (1987). (A)
\end{itemize}
The corresponding force on an atom by a perfectly conducting wedge
was was only computed a decade later.
\begin{itemize}
\yitem ``Casimir-Polder Effect for a Perfectly Conducting Wedge,''
I. Brevik,  M. Lygren, and V. N. Marachevsky, Ann.\ Phys.\ (N.Y.) {\bf267},
134--142 (1998). (I)
\end{itemize} 
An earlier experiment by Hinds' group had seen the expected scaling of
the van der Waals deflection of high $n$ Rydberg atoms:
\begin{itemize}
\yitem ``Measuring the van der Waals Forces between a Rydberg Atom and a 
Metallic Surface,'' A. Anderson, S. Haroche, E. A. Hinds, W. Jhe, and 
D. Meschede, Phys.\ Rev.\ A {\bf37}, 3594--3597 (1988). (I)
\end{itemize}
There were precursors to these investigations, for example,
\begin{itemize}
 \yitem ``Interaction between a Neutral Atomic or Molecular Beam and a Conducting Surface,''
D. Raskin and P. Kusch, Phys.\ Rev.\ {\bf179}, 712--721 (1969). (I)
\end{itemize}

More recently, however, the force between a Bose-Einstein condensate
of ${}^{87}$Rb 
atoms and a dielectric substrate was measured, even at different temperatures,
by Eric Cornell's group:
\begin{itemize}
 \yitem ``Measurement of the Casimir-Polder Force Through Center-of-Mass 
Oscillations of a Bose-Einstein Condensate,''
D. M. Harber, J. M. Obrecht, J. M. McGuirk, and E. A. Cornell,
Phys.\ Rev.\ A {\bf72}, 033610 (6 pages) (2005). (A)
\yitem ``Measurement of the Temperature Dependence of the Casimir-Polder 
Force,'' J. M. Obrecht, R. J. Wild, M. Antezza, L. P. Pitaevskii, 
S. Stringari, and E. A. Cornell, Phys.\ Rev.\ Lett.\ {\bf98}, 063201 
(4 pages) (2007). (I) 
\end{itemize}

\section{Developments in van der Waals--Casimir-Polder forces}\label{sec:devcp}
Van der Waals forces can be thought of as arising from two-photon exchange:
\begin{itemize}
\yitem ``The Dispersion Theory of Dispersion Forces,'' 
G. Feinberg, J. Sucher, and C. -K. Au, Phys.\ Rept.\ {\bf 180}, 83--157 (1989).
(I)
\end{itemize}
In this paper, the authors rederive the Casimir-Polder (\ref{cp}) and 
London forces (\ref{london}), and do so using a scattering matrix formalism,
which has recently become popular (see Sec.~\ref{secexact}).  
One-photon exchange processes yield the
Coulomb interaction, while two-photon processes lead to a potential between
two atoms, labeled $A$ and $B$, of the form 
\be
V_{\rm CP}=-\frac{D}{r^7}\hbar c,
\ee
where in general (for isotropic atoms)
\be
D=\frac{23}{4\pi}\left(\alpha_E^A\alpha_E^B+\alpha_M^A\alpha_M^B\right)
-\frac{7}{4\pi}\left(\alpha_E^A\alpha_M^B+\alpha_M^A\alpha_E^B\right),
\label{gencp}
\ee 
in terms of the electric and magnetic polarizabilities $\alpha_{E,M}$.
This result was first obtained by Feinberg and Sucher twenty years earlier,
using dispersion theory.
\begin{itemize}
\yitem ``General Form of the Retarded van der Waals Potential,''
G. Feinberg and J. Sucher, J. Chem.\ Phys.\ {\bf48}, 3333--3334 (1968). (I)
\end{itemize}
It was then reproduced on a more field theoretic
basis by Boyer.
\begin{itemize}
\yitem ``Recalculations of Long-Range van der Waals Potentials,'' T. H.
Boyer, Phys.\ Rev.\ {\bf180}, 19--24 (1969). (I)
\end{itemize}
In their earlier paper Feinberg and Sucher noted that while the magnetic
polarizability is usually negligible, it is not true for the hydrogen atom,
where the coefficient of the $1/r^7$ is about 750 times that given by Casimir
and Polder, although in that case the retarded interaction is only valid for
$r>21$ cm, making ``the matter somewhat academic.'' 
This result (\ref{gencp}) reduces to the Casimir-Polder result if $\alpha_M=0$.  In general,
this allows for repulsive interactions, although such a situation is
hard to realize in practice (see Sec.~\ref{secrep}). Feinberg and Sucher
also note in their 1968 paper that ``it would be very interesting, and a
new confirmation of quantum electrodynamics, if some way could be found
to detect such a repulsion.''  This is an equally relevant comment
today.  See also
\begin{itemize}
\yitem ``Quantum Zero-Point Energy and Long-Range Forces,'' T. H. Boyer,
Ann.\ Phys.\ {\bf 56}, 474--503 (1970). (I)
\yitem ``Simple Derivation of the Asymptotic Casimir Interaction of a Pair 
of Finite Systems,'' L. Spruch, J. F. Babb, and F. Zhou, Phys.\ Rev.\ A 
{\bf49}, 2476--2482 (1994). (I)
\yitem ``Long Range Electromagnetic Effects Involving Neutral Systems and 
Effective Field Theory,'' B. R. Holstein, Phys.\ Rev.\ D {\bf 78}, 013001
(11 pages) (2008). (I)
\end{itemize}

There are a great many developments in computing CP forces for atomic
systems.  For example, see:
\begin{itemize}
\yitem ``Casimir-Polder Forces on Excited Atoms in the Strong Atom-Field 
Coupling Regime,'' S. Y. Buhmann and D.-G. Welsch, Phys.\ Rev.\ A {\bf 77},
012110 (16 pages) (2008). (A)
\end{itemize}

The analogous retarded Casimir interaction between an electron and a dielectric
wall was given in the following.
\begin{itemize}
\yitem ``Retarded Casimir Interaction in the Asymptotic Domain of an Electron 
and a Dielectric Wall,'' Y. Tikochinsky and L. Spruch, Phys. Rev. A {\bf48}, 
4223--4235 (1993). (I)
\end{itemize}
\section{Field theory of Casimir forces and QCD}

Symanzik thirty years ago provided a general renormalized field-theoretic 
basis to the Casimir effect: 
\begin{itemize}
\yitem ``Schr\"odinger Representation and Casimir Effect in Renormalizable 
Quantum Field Theory,''
K. Symanzik, Nucl.\ Phys.\ B {\bf190}, 1--44 (1981). (A)
\end{itemize}
He noted there, perhaps before it was generally recognized, that
``the Casimir potential between disjoint surfaces is always well defined.
That for a single surface, e.g. a sphere, in general is not.''  There are
still issues to be resolved in Casimir self-energies.
\begin{itemize}
\yitem \label{milton11}
``Local and Global Casimir Energies: Divergences, Renormalization, 
and the Coupling to Gravity,'' K. A. Milton,
 arXiv:1005.0031 (53 pages), 
invited review paper to appear in a Lecture Notes in 
Physics Volume on Casimir physics edited by Diego Dalvit, Peter Milonni, 
David Roberts, and Felipe da Rosa. (A)
\end{itemize}
See Sec.~\ref{secse}.

The gauge theory of strong interactions, quantum chromodynamics (QCD), has
still not been completely solved.  A simple model that encompasses some of
the expected features of the full theory is the MIT bag model.  In this
model, there is an important parameter that is supposed to represent
the effects of zero-point fluctuations of the quark and gluon fields within
the confining bag.  Crude estimates of these effects were computed some
thirty years ago:
\begin{itemize}
\yitem ``Zero-point Energy in Bag Models,'' K. A. Milton, Phys.\ Rev.\
D {\bf 22}, 1441--1443 (1980). (I)
\yitem ``Zero-point Energy of Confined Fermions,'' K. A. Milton, Phys.\
Rev.\ D {\bf 22}, 1444--1451 (1980). (A)
\end{itemize}
Other parameters of QCD, of phenomenological significance, could
also be estimated in a bag-model context:
\begin{itemize}
\yitem ``Quark and Gluon Condensates in a Bag Model of the Vacuum,''
K. A. Milton, Phys.\ Lett.\ B {\bf104}, 49--54 (1981). (A)
\end{itemize}
In this connection, the important early paper of Bender and Hays must be
mentioned:
\begin{itemize}
 \yitem ``Zero-Point Energy of Fields in a Finite Volume,''
C. M. Bender and P. Hays,
Phys.\ Rev.\ D {\bf14}, 2622 (1976). (I)
\end{itemize}

About the same time, Feinberg and Sucher suggested that there might be
a strong van der Waals force between hadrons, which in turn could be a
 residual effect of QCD within the hadrons:
\begin{itemize}
\yitem  ``Is There a Strong van der Waals Force Between Hadrons?''
G.~Feinberg, J.~Sucher,
  Phys.\ Rev.\ D {\bf 20}, 1717--1735 (1979). (I)
\end{itemize}
The authors noted that such long-range forces are subject to severe
phenomenological constaints, and for example are absent in the bag model.
See also the following:
\begin{itemize}
\yitem  ``A Multipole Expansion and the Casimir-Polder
Effect in Quantum Chromodynamics,''
G.~Bhanot, W.~Fischler, and S.~Rudaz,
  Nucl.\ Phys.\  B {\bf 155}, 208--236 (1979).
\end{itemize}
Casimir-Polder forces in QCD were considered more
recently in
\begin{itemize}
 \yitem ``Long-Range Forces of QCD,'' H. Fujii and D. Kharzeev,
Phys.\ Rev.\ D {\bf 60}, 114039 (12 pages) (1999). (A) 
\end{itemize}
An analysis of Casimir energies relevant to both QCD and
cosmology was given in the important paper by Blau, Visser,
and Wipf:
\begin{itemize}
 \yitem ``Zeta Functions and the Casimir Energy,''
S. K. Blau, M. Visser, and A. Wipf, Nucl.\ Phys.\ B
{\bf 310}, 163--180 (1988). (A)
\end{itemize}

\section{Exact Casimir Force from C-P Forces}
We can reverse the derivation of the Casimir-Polder force from
the Lifshitz forces between dilute dielectrics referred to above, in
Sec.~\ref{sec:ce},  to
calculate quantum vacuum forces between macroscopic bodies made up
of polarizable molecules.  In many cases these forces can be given
closed form through summing the Casimir-Polder potentials (\ref{cp}):
\begin{itemize}
\yitem ``Exact Results for Casimir Interactions between Dielectric Bodies: 
The Weak-Coupling or van der Waals Limit,''
K. A. Milton, P. Parashar, and J. Wagner,
Phys.\ Rev.\ Lett.\ {\bf101}, 160402 (4 pages) (2008). (I) 
\end{itemize}
For example, one can calculate exact forces between dilute
bodies of various shapes,
and can exhibit a Casimir torque between a semi-infinite  dielectric slab and a
dielectric rectangular block: For a fixed distance between the center of
mass of the block and the slab, 
the equilibrium configuration of the block ranges from that in 
which the shortest side faces the plane, for large distances, 
to one in which the center of mass of the block is directly above the
point of contact when the bodies just touch.  This is a ``tidal'' effect
hinging on the fact that the CP force falls off with distance.
  
Such pairwise summations of CP forces were first carried out to calculate
self energies.  For example, the self energy of a  ball 
of radius $a$ made of a dilute
dielectric $|\varepsilon-1|\ll1$, was evaluated by summing Casimir-Polder
energies.
\begin{itemize}
\yitem ``Observability of the Bulk Casimir Effect: Can the Dynamical 
Casimir Effect be Relevant to Sonoluminescence?''
K. A. Milton and Y. J. Ng,
Phys.\ Rev.\ E {\bf57}, 5504--5510 (1998). (I)
\end{itemize}
The resulting self-energy was found to be
\be
E_{\rm dilute\, sphere}=\frac{23}{2536\pi a}(\varepsilon-1)^2,
\ee
which was later verified by a full field-theoretic calculation.
\begin{itemize}
\yitem ``Identity of the van der Waals Force and the Casimir Effect and 
the Irrelevance of These Phenomena to Sonoluminescence,''
I. Brevik, V. N. Marachevsky, and K. A. Milton,
Phys.\ Rev.\ Lett.\ {\bf82}, 3948--3951 (1999). (A)
\end{itemize}
The corresponding energy for a dilute dielectric circular cylinder
is zero, verified both by summing Casimir-Polder energies, and a full
calculation of the Casimir energy:
\begin{itemize}
\yitem ``Mode-by-Mode Summation for the Zero Point Electromagnetic Energy of 
an Infinite Cylinder,''
K. A. Milton, A. V. Nesterenko, and V. V. Nesterenko,
Phys.\ Rev.\ D {\bf59}, 105009 (9 pages) (1999). (I)
\yitem ``Casimir Energy for a Dielectric Cylinder,'' I. Cavero-Pel\'aez
and K. A. Milton, Ann.\ Phys.\ (N.Y.) {\bf320}, 108--134 (2005). (A)
\yitem ``Casimir Energy for a Purely Dielectric Cylinder by the Mode
Summation Method,'' A. Romeo and K. A. Milton, Phys.\ Lett.\ B {\bf 621},
309--317 (2005). (A)
\end{itemize}
Vanishing of weakly coupled cylindrical Casimir energies is a universal
feature, which is not yet well understood.

Actually, there is a much more extensive literature calculating Casimir forces
between dilute  bodies based on the nonretarded van der Waals interaction
going like $r^{-6}$.  Much of this is summarized in the book by Parsegian.
\begin{itemize}
\yitem {\bf Van der Waals Forces: A Handbook for Biologists, Chemists,
Engineers, and Physicists,} V. Adrian Parsegian (Cambridge University
Press, Cambridge, 2007). (I)
\end{itemize}
Nonretarded van der Waals forces have importance in biological systems, 
but the use of the unretarded
form of the force is typically restricted to distances below 10 nm.
Special mention should be made of the following paper.
\begin{itemize}
\yitem ``On the Macroscopic Theory of van der Waals Forces,''
N. G. Van Kampen, B. R. A. Nijboer, and K. Schram,
Phys.\ Lett.\ A {\bf26}, 307--308 (1968). (I)
\end{itemize}
They rederived the Lifshitz formula in the non-retarded regime; we note
the following precusor and follow-up:
\begin{itemize}
\yitem ``Microscopic Derivation of Macroscopic van der Waals Forces,''
M. J. Renne and B. R. A. Nijboer, Chem.\ Phys.\ Lett. {\bf 1}, 317--320 (1967).
(I)
\yitem ``On the Macroscopic Theory of Retarded van der Waals Forces,''
K. Schram, Phys.\ Lett.\ A {\bf 43}, 282--284 (1973). (I)
\end{itemize}

Broad strokes of how Casimir phenomena could be relevant in chemistry
are given in the following.
\begin{itemize}
\yitem ``Casimir Chemistry,'' D. P. Sheehan, J. Chem.\ Phys.\ {\bf 131},
104706 (11 pages) (2009). (E)
\end{itemize}

\section{Theoretical controversies and experimental constraints}\label{sec:con}
Some years ago Bordag suggested that the theory of Casimir-Polder forces
might be subject to modification, if the conducting boundaries are ``thin.''
The resulting force is modified from that of the standard theory by about
13\%, which is consistent with the experimental error in the Sukenik 
experiment, Ref.~[\ref{sukenik}].
\begin{itemize}
\yitem 
``Reconsidering the Quantization of Electrodynamics with Boundary Conditions 
and Some Measurable Consequences,''
M.~Bordag,    Phys.\ Rev.\ D {\bf 70}, 085010 (11 pages) (2004).  (I) 
\end{itemize}
This result remains controversial, since no sign of this
effect appears in the calculation based on field-strength tensors
found in the Schwinger-DeRaad-Milton
paper,  Ref.~[\ref{sdm}].

A much more heated controversy has arisen concerning the temperature
dependence of the Casimir effect for metals.  As mentioned above,
in Sec.~\ref{sec:ce},
the prescription of extracting the temperature dependence for perfect
metals used by Lifshitz, made most explicit in Ref.~[\ref{sdm}], was
justified later by a desire not to conflict with the third law of thermodynamics.
But Bostr\"om and Sernelius argued that this could not be correct.
\begin{itemize}
\yitem ``Thermal Effects on the Casimir Force in the
0.1--5 $\mu$m Range,''  M. Bostr\"om and Bo E. Sernelius,
Phys.\ Rev.\ Lett.\ {\bf84}, 4757--4760 (2000). (I)
\end{itemize}
They showed that the TE zero-frequency
mode should be excluded for metals, just as it is for dielectrics.
This has led to some intense discussion, partly to do with the
thermodynamic inconsistency. 
\begin{itemize}
 \yitem \label{bezerra02}
``Correlation of Energy and Free Energy for the Thermal Casimir 
Force between Real Metals.'' V. B. Bezerra, G. L. Klimchitskaya, and 
V. M. Mostepanenko, Phys.\ Rev.\ A {\bf66}, 062112 (13 pages) (2002). (A)
\yitem \label{bezerra04}
``Violation of the Nernst Heat Theorem in the Theory of the Thermal 
Casimir Force between Drude Metals,'' V. B. Bezerra, G. L. Klimchitskaya, 
V. M. Mostepanenko, and C. Romero, Phys.\ Rev.\ A {\bf69}, 022119 (9 pages) (2004). (A)
\end{itemize}
 In fact, it has now been shown that
for real metals (not fictional ideal perfect crystal lattices) the third law is satisfied,
in that the entropy vanishes at zero temperature:
\begin{itemize}
\yitem
``Analytical and Numerical Verification of the Nernst Theorem for Metals,''
J. S. H\o ye, I. Brevik, S. A. Ellingsen, and J. B. Aarseth,
Phys.\ Rev.\ E {\bf75}, 051127 (8 pages) (2007). (A)
\yitem ``Analytical and Numerical Demonstration of How the Drude Dispersive 
Model Satisfies Nernst's Theorem for the Casimir Entropy,''
I. Brevik, S. A. Ellingsen, J. S. H\o ye, and  K. A. Milton, J. Phys.\ A: 
Math.\ Theor.\ {\bf41}, 164017 (10 pages) (2008). (I)
\end{itemize}
Now there is general agreement that the modified theory is correct, although
the authors of Refs.~[\ref{bezerra02}] and [\ref{bezerra04}] would disagree.
This means that at low temperature there is a linear temperature correction
term, and at high temperature the usual linear temperature dependence is
reduced by a factor of 2.  If the metal is otherwise ideal, the pressure given
by the modified theory is [$\zeta(3)=1.20206$ is the Riemann zeta function at 3]
\begin{subequations}
\begin{eqnarray}
P&=&-\frac{\pi^2\hbar c}{240 a^4}\left[1+\frac{16}3\left(\frac{a k T}{\hbar c}
\right)^4\right]+\frac{\zeta(3)}{8\pi a^3} k T,\quad kT a/\hbar c\ll 1,
\label{lowt}
\\
P&=&-\frac{\zeta(3)}{8\pi a^3} k T,\quad kT a/\hbar c\gg 1,\label{hight}
\end{eqnarray}
\end{subequations}
while the linear term in the low-temperature limit is not present in
the conventional theory, and the linear term in the high-temperature limit
is twice as large in the conventional theory.  
The reduction of the high-temperature effect seen in Eq.~(\ref{hight}) 
was found in independent theoretical approaches:
\begin{itemize}
\yitem ``Casimir Force between Two Ideal-Conductor Walls Revisited,''
B. Jancovici and L. \v{S}amaj, Europhys.\ Lett.\ {\bf72}, 35--41 (2005). (I)
\yitem ``The Casimir Force at High Temperature,'' P. R. Buenzli and 
Ph. A. Martin, Europhys.\ Lett.\ {\bf72}, 42--48 (2005). (I)
\end{itemize}
The thermodynamic problem
associated with Eq.~(\ref{lowt}) is that the corresponding free energy is
\begin{equation}
F\approx F_0+\frac{\zeta(3)}{16\pi a^2} k T,\quad P=-\left(\frac{\partial
F}{\partial a}\right)_T,\label{convfe}
\end{equation}
at low temperature, so the entropy/area does not vanish at low $T$:
\begin{equation}
S=-\left(\frac{\partial F}{\partial T}\right)_V=-\frac{\zeta(3)}{16\pi a^2}
 k.
\end{equation}
But, in fact, it turns out that,
 because for real metals, $\zeta^2\varepsilon(\zeta)
\to 0$ as the frequency $\zeta\to0$,  for very low temperatures
the free energy behaves not as Eq.~(\ref{convfe}) but as
\begin{equation}
F=F_0+\gamma T^2,\quad kT\ll 20 \mbox{mK},
\end{equation}
where $\gamma$ is a constant depending on the resistivity of the metal. 
[$\gamma(T=0)\ne0$ for a real metal; $\gamma(T=0)=0$ only for a perfect
ideal lattice metal.]  Thus
the entropy vanishes at $T=0$, and there is no thermodynamic inconsistency.
These results agree fairly closely with those obtained through
 exact integration of the Lifshitz
formula using optical data for real metals, such as gold.
The prediction (\ref{lowt})  is claimed by the authors
of Refs.~[\ref{decca05}], [\ref{decca07}], [\ref{decca07a}]
 to be insconsistent with
their experiments, but this is not persuasive to
many observers, since the thermal effect at room temperatures for experiments
conducted at the 100 nm distance range is only a few percent.  The 
prediction (\ref{hight}) has
now been verified by a new experiment by Lamoreaux's group in the
micrometer range:
\begin{itemize}
\yitem ``Observation of the Thermal Casimir Force,''
A. O. Sushkov, W. J. Kim, D. A. R. Dalvit, and S. K. Lamoreaux,
Nature Physics {\bf7}, 230--233 (2011). (I)
\end{itemize}
See also the commentary on this paper:
\begin{itemize}
 \yitem  ``The Casimir Force: Feeling the Heat,'' K. A. Milton,
Nature Physics {\bf7}, 190--191 (2011). (E)
\end{itemize}
Not surprisingly, Mostepanenko and collaborators do not accept this
result, and argue that surface imperfections in the large spheres
used in this experiment render the experiment inconclusive.
\begin{itemize}
 \yitem ``Impact of Surface Imperfections on the Casimir Force 
for Lenses of Centimeter-Size Curvature Radii,''
V. B. Bezerra, G. L. Klimchitskaya, U. Mohideen, V. M. Mostepanenko, 
and C. Romero, Phys.\ Rev.\ B {\bf83}, 075417 (13 pages) (2011). (I)
\end{itemize}
 
There is also a controversy about the thermal Casimir force between semiconductors.
In this case, the discontinuity involves the zero-frequency TM mode.  The problem
was identified in 2005.
\begin{itemize}
 \yitem ``Thermal Quantum Field Theory and the Casimir Interaction between Dielectrics,''
B. Geyer, G. L. Klimchitskaya, and V. M. Mostepanenko,
Phys.\ Rev.\ D {\bf72}, 085009 (20 pages) (2005). (A) 
\end{itemize}
The difficulty was elaborated in further papers by the same authors.
A different perspective was given by in the following reference.
\begin{itemize}
 \yitem ``Temperature Correction to Casimir-Lifshitz Free Energy at Low Temperatures: 
Semiconductors,'' S. \AA. Ellingsen, I. Brevik, J. S. H\o ye, and K. A. Milton,
Phys.\ Rev.\ E {\bf78}, 021117 (11 pages) (2008). (A)
\end{itemize}
A related experimental anomaly was discovered when the carrier concentration
of a semiconductor was changed by illumination by a laser.  The data indicated
that the conductivity must be excluded for a low carrier concentration, but
included in the case of high carrier concentration.  Inclusion of the
conductivity (which is nonzero in all cases) in both situations led to disagreement
with the experimental results. 
\begin{itemize}
 \yitem ``Control of the Casimir Force by the Modification of 
Dielectric Properties with Light,''
F. Chen, G. L. Klimchitskaya, V. M. Mostepanenko, and U. Mohideen,
Phys.\ Rev.\ B {\bf76}, 035338 (15 pages) (2007). (I)
\end{itemize}
These issues have yet to be satisfactorily resolved.

On the other hand, there seems to be no controversy about the
thermal Casimir-Polder force between an atom and a metallic surface. 
However, there appears to be a conflict between theory and experiment,
and with the Nernst heat theorem for the CP force between an atom and
a dielectric if the dc conductivity of the latter in included.
\begin{itemize}
 \yitem ``Conductivity of Dielectric and Thermal Atom-Wall Interaction,''
G. L. Klimchitskaya and V. M. Mostepanenko,
J. Phys.\ A: Math.\ Theor.\ {\bf41},  312002  (8 pages) (2008). (I)
\yitem ``Thermal Casimir-Polder Force between an Atom and a Dielectric Plate: 
Thermodynamics and Experiment,''
G. L. Klimchitskaya, U. Mohideen, and V. M. Mostepanenko,
 J. Phys.\ A: Math.\ Theor.\ {\bf41}, 432001
(9 pages) (2009). (I)
\end{itemize}
As noted above, this problem of the effect of small conductivity is
still under investigation.
\begin{itemize}
 \yitem ``Thermal Lifshitz Force between an Atom and a Conductor with a 
Small Density of Carriers,'' L. P. Pitaevskii,
Phys.\ Rev.\ Lett.\ {\bf 101}, 163202 (4 pages) (2008). (I)
\end{itemize}

Interestingly,
the CP potential may be temperature-independent even when the background
contains a large number of thermal photons.
\begin{itemize}
\yitem ``Temperature-Independent Casimir-Polder Forces Despite Large 
Thermal Photon Numbers,'' 
S. \AA. Ellingsen, S. Y. Buhmann, and S. Scheel,
Phys.\ Rev.\ Lett.\ {\bf104}, 223003 (4 pages) (2010). (I)
\end{itemize}
The classical regime of linear temperature dependence (\ref{cpt}) is never
reached.   There is also recent interesting work on out-of-equilibrium
interatomic forces:
\begin{itemize} 
\yitem ``Casimir-Lifshitz Force out of Thermal Equilibrium,''
M. Antezza, L. P. Pitaevskii, S. Stringari, and V. B. Svetovoy,
Phys.\ Rev.\ A {\bf77}, 022901  (22 pages) (2008). (A)
\yitem ``Nonequilibrium Forces between Neutral Atoms Mediated by a Quantum 
Field,'' R. O. Behunin and B.-L. Hu, Phys.\  Rev.\ A {\bf82}, 022507 
(12 pages) (2010). (I)
\yitem ``Dynamics of Thermal Casimir-Polder Forces on Polar Molecules,''
S. \AA. Ellingsen, S. Y. Buhmann, and S. Scheel, Phys.\ Rev.\ A {\bf 79},
052903 (6 pages) (2009). (I)
\end{itemize}
The latter reference shows that out of equilibrium, nonmonotonic 
(repulsive) forces can occur.

\section{Nanotechnological applications}\label{sec:nano}
The first experimental papers that demonstrated that quantum
vacuum forces could be useful in micromachinery appeared ten years ago:
\begin{itemize}
\yitem ``Stiction, Adhesion Energy, and the Casimir Effect in Micromechanical Systems,''
E. Buks and M. L. Roukes,
Phys.\ Rev.\ B {\bf63}, 033402 (4 pages) (2001). (I)
\yitem ``Metastability and the Casimir Effect in Micromechanical Systems,''
E. Buks and  M. L. Roukes,
Europhys.\ Lett.\ {\bf54},  220--226 (2001). (I)
\yitem ``Quantum Mechanical Actuation of Microelectromechanical Systems 
by the Casimir Force,''
H. B. Chan, V. A. Aksyuk, R. N. Kleiman, D. J. Bishop, and F. Capasso, 
 Science {\bf 291}, 1941--1944 (2001). (I)
\yitem ``Nonlinear Micromechanical Casimir Oscillator,''
H. B. Chan, V. A. Aksyuk, R. N. Kleiman, D. J. Bishop, and F. Capasso,
Phys.\ Rev.\ Lett.\ {\bf87}, 211801 (4 pages) (2001). (I)
\end{itemize}
The first two references here find forces that do not agree with the
simple Casimir theory.

There have been many theoretical papers proposing micro-machinery and nano-machinery
actuated by Casimir-type forces, for example, the proposal for a noncontact
rack and pinion:
\begin{itemize}
\yitem ``Noncontact Rack and Pinion Powered by the Lateral Casimir Force,''
A. Ashourvan, M. Miri, and R. Golestanian,
Phys.\ Rev.\ Lett.\ {\bf98}, 140801 (4 pages) (2007). (I)
\end{itemize}
The following papers suggested the development of non-contact gears:
\begin{itemize}
\yitem ``Exploring the Quantum Vacuum with Cylinders,''
F. C. Lombardo, F. D. Mazzitelli, and P. I. Villar,
J. Phys.\ A: Math.\ Theor.\ {\bf41},  164009 (10 pages) (2008). (I)
\yitem ``Noncontact Gears. II. Casimir Torque Between Concentric Corrugated 
Cylinders for the Scalar Case,''
I. Cavero-Pel\'aez,  K. A. Milton, P. Parashar, and K. V. Shajesh,
Phys.\ Rev.\ D {\bf78}, 065018 (7 pages) (2008). (A)
\end{itemize}
For a perspective from the Paris group, see the following.
\begin{itemize}
\yitem ``The Casimir Effect in the Nanoworld,''
C. Genet, A. Lambrecht, and S. Reynaud,
Eur.\ Phys.\ J. Special Topics {\bf160}, 183 (8 pages) (2008)
[arXiv:0809.0254]. (I)
\end{itemize}
\section{Casimir friction}
Lateral as opposed to normal forces between plates have been considered
in a variety of contexts in recent years.  For example, if the plates are
corrugated, there should be a lateral (sideways) force attempting to
bring the peaks of the lower surface closest to the troughs in the upper
surface.  This was first observed by Mohideen's group:
\begin{itemize}
\yitem ``Demonstration of the Lateral Casimir Force,''
F. Chen, U. Mohideen, G. L. Klimchitskaya, and V. M. Mostepanenko,
Phys. Rev. Lett. {\bf88}, 101801 (4 pages) (2002). (I)
\end{itemize}
Comparison with the best theory at the time, the proximity-force approximation,
was not very satisfactory:
\begin{itemize}
\yitem ``Experimental and Theoretical Investigation of the Lateral Casimir 
Force Between Corrugated Surfaces,''
F. Chen, U. Mohideen, G. L. Klimchitskaya, and V. M. Mostepanenko,
Phys. Rev. A {\bf66}, 032113 (11 pages) (2002). (I)
\yitem ``Lateral Casimir Force Beyond the Proximity Force Approximation: 
A Nontrivial Interplay Between Geometry and Quantum Vacuum,''
R. B. Rodrigues, P. A. Maia Neto, A. Lambrecht, and S. Reynaud,
Phys.\ Rev.\ A {\bf75}, 062108 (10 pages) (2007). (I)
\end{itemize}
The definitive observation, and comparison with the multiple-scattering
theory, was given in 2009:
\begin{itemize}
\yitem ``Lateral Casimir Force Between Sinusoidally Corrugated Surfaces: 
Asymmetric Profiles, Deviations from the Proximity Force Approximation, and 
Comparison with Exact Theory,''
H.-C. Chiu, G. L. Klimchitskaya, V. N. Marachevsky, V. M. Mostepanenko, and 
U. Mohideen,
Phys.\ Rev.\ B {\bf81}, 115417 (20 pages)(2010). (I)
\end{itemize}
  This is relevant to
nanotechnological applications. (See Sec.~\ref{sec:nano}.)
However, with deeper trenches on the surfaces, the deviation from
the proximity force approximation is more pronounced.
\begin{itemize}
\yitem ``Measurement of the Casimir Force between a Gold Sphere and a 
Silicon Surface with Nanoscale Trench Arrays,''
H. B. Chan, Y. Bao, J. Zou, R. A. Cirelli, F. Klemens, W. M. Mansfield, 
and C. S. Pai, Phys.\ Rev.\ Lett.\ {\bf101}, 030401 (4 pages) (2008). (I)
\end{itemize}
The results agree with theoretical calculations:
\begin{itemize}
 \yitem ``Casimir Interaction of Dielectric Gratings,''
A. Lambrecht and V. N. Marachevsky,
Phys.\ Rev.\ Lett.\ {\bf101}, 160403 (4 pages) (2008). (I) 
\end{itemize}

But can there be a transverse force in the absence of corrugations?
Sometime ago, researchers started considering the transverse force between plane
surfaces in relative parallel motion:
\begin{itemize}
\yitem ``On the Contribution of Macroscopic van der Waals Interactions to 
Frictional Force,''
E. V. Teodorovich, Proc.\ R. Soc.\ London, Ser. A, {\bf362}, 71--77
(1978). (I)
\yitem  ``Van der Waals' Friction,'' L. S. Levitov, Europhys. Lett. 
{\bf8}, 499--504 (1989). (I) 
\yitem ``The `Friction' of Vacuum, and Other Fluctuation-Induced Forces,'' 
M. Kardar and R. Golestanian, Rev.\ Mod.\ Phys.\ {\bf71}, 1233--1245 (1999).
(I)
\end{itemize}
Some authors find no friction at all:
\begin{itemize}
\yitem ``No Quantum Friction Between Uniformly Moving Plates,''
T. G. Philbin and U. Leonhardt, New J. Phys.\ {\bf11}, 
033035 (18 pages) (2009). (I)
\end{itemize}
A useful overview is
\begin{itemize}
\yitem  ``Quantum Friction---Fact or Fiction?''
J. B. Pendry,  New J. Phys.\ {\bf12},  033028 (7 pages) (2010). (I)
\end{itemize}
The extensive work of Dedkov and Kyasov should be cited:
\begin{itemize}
 \yitem ``Fluctuation Electromagnetic Slowing Down and Heating of a Small Neutral 
Particle Moving in the Field of Equilibrium Background Radiation,''
 G. V. Dedkov and A. A. Kyasov, Phys.\ Lett.\ A {\bf339} 212--216 (2005). (I)
\end{itemize}
See also earlier papers cited therein, as well as the following.
\begin{itemize}
 \yitem ``Casimir-Polder Forces on Moving Atoms,''
S. Scheel and S. Y. Buhmann, 
Phys.\ Rev.\ A {\bf80}, 042902 (11 pages) (2009). (A)
\end{itemize}

A statistical-mechanical treatment of the Casimir friction problem, based
on dielectric particles moving with constant relative velocity, is given in
\begin{itemize}
\yitem \label{brevikf}
``Casimir Friction Force and Energy Dissipation for Moving Harmonic 
Oscillators,'' J. S. H\o ye and I. Brevik, Europhys.\ Lett.\ {\bf 91}, 60003
(5 pages) (2010). (I)
\end{itemize}
They conclude that the friction vanishes at zero temperature, but is nonzero
at nonzero temperature.
A different perspective on this is given by Barton:
\begin{itemize}
\yitem ``On van der Waals Friction: I. Between Two Atoms,''
G. Barton,
New J. Phys.\ {\bf12},  113044 (10 pages) (2010). (I)
\yitem ``On van der Waals Friction. II: Between Atom and Half-Space,''
G. Barton,
New J. Phys.\ {\bf12}, 113045 (13 pages) (2010). (I)
\end{itemize}
For example, Barton finds friction at all temperatures, a difference
with the result found in Ref.~[\ref{brevikf}] that he 
attributes to a different choice of initial conditions.
However, the authors of that reference demonstrate that both
approaches are in fact equivalent.
\begin{itemize}
\yitem ``Casimir Friction in Terms of Moving Harmonic Oscillators: 
Equivalence Between Two Different Formulations,'' J. S. H\o ye
and I. Brevik, arXiv:1101.1241. (I)
\end{itemize}
\section{Recent theoretical developments}\label{secexact}
In the last few years, great progress has been made in calculating
forces between objects of arbitrary shape. Previously, deviations
of forces from that of plane surfaces had been calculated
using the proximity force approximation [\ref{ref:pfa}].  The
associated errors of this approximation were unknown.
One approach for transcending such limitations was the
``world-line'' method developed by Gies and collaborators:
\begin{itemize}
 \yitem ``Worldline Algorithms for Casimir Configurations,''
H. Gies and K. Klingmuller, Phys.\ Rev.\ D {\bf74}, 045002 (12 pages) (2006). (A)
\yitem ``Geothermal Casimir Phenomena,''
K. Klingmuller and H. Gies, J. Phys.\ A {\bf41}, 164042 (8 pages) (2008). (A)
\yitem ``Interplay between Geometry and Temperature for Inclined Casimir
Plates.'' A. Weber and H. Gies,
 Phys.\ Rev.\ D {\bf80}, 065033 (17 pages) (2009). (A)
\yitem ``Nonmonotonic Thermal Casimir Force from Geometry-Temperature
Interplay,'' A. Weber and H. Gies, Phys.\ Rev.\ Lett.\ {\bf105}, 040403 (4 pages) (2010).
(A)
\yitem ``Geometry-Temperature Interplay in the Casimir Effect,''
H. Gies and A. Weber, Int.\ J. Mod.\ Phys.\ A {\bf25}, 2279--2292 (2010). (A)
\end{itemize}
Unfortunately, no one has yet succeeded in applying this method to other
than perfect Dirichlet boundary conditions.

 Much of the recent advance has to do
with the application of multiple scattering techniques, which of course
have a long history.  These developments early on included
cylindrical calculations, which might be amenable to experimentation.
\begin{itemize}
 \yitem ``Exact Zero-Point Interaction Energy Between Cylinders,''
F. D. Mazzitelli, D. A. R. Dalvit, and F. C. Lombardo,
New J. Phys.\ {\bf 8}, 240 (21 pages) (2006). (A)
\yitem ``Exact Casimir Interaction between Eccentric Cylinders,''
D. A. R. Dalvit, F. C. Lombardo, F. D. Mazzitelli, and
R. Onofrio, Phys.\ Rev.\ A {\bf 74}, 020101 (4 pages) (2006). (A)
\end{itemize}
More general developments appeared in the following papers.
\begin{itemize}
 \yitem ``Scalar Casimir Effect between Dirichlet Spheres or a Plate and a
  Sphere,'' A. Bulgac, P. Magierski, and A. Wirzba,
Phys.\ Rev.\ D {\bf73}, 025007 (14 pages) (2006). (A)
\yitem ``Casimir Interaction between a Plate and a Cylinder,''
T.~Emig, R.~L. Jaffe, M.~Kardar, and A.~Scardicchio,
 Phys.\ Rev.\ Lett.\ {\bf96}, 08040 (4 pages) (2006). (A)
\yitem ``Opposites Attract---A Theorem about the Casimir Force,''
O. Kenneth and I. Klich, Phys.\ Rev.\ Lett.\ {\bf97}, 160401 (4 pages) (2006). (I)
\yitem \label{emigjsm}
``Fluctuation-Induced Quantum Interactions between Compact Objects and
  a Plane Mirror,'' T.~Emig, J. Stat.\ Mech.--Th.\ Exp.\
  P04007 (33 pages) (2008). (A)
\end{itemize}
In particular, the sphere-plate configuration used in experiments, was
analyzed in the last reference and in the following.
\begin{itemize}
 \yitem ``Scattering Theory Approach to Electrodynamic Casimir Forces,''
S. J. Rahi, T. Emig, N. Graham, R. L. Jaffe, and M.  Kardar,
Phys.\ Rev.\ D {\bf80}, 085021 (27 pages) (2009). (A)
\yitem \label{mnlr}
``Casimir Energy between a Plane and a Sphere in Electromagnetic
  Vacuum,'' P. A. Maia Neto, A. Lambrecht, and S. Reynaud,
Phys.\ Rev.\ A {\bf78},  012115 (4 pages) (2008). (A)
\end{itemize}
Work continues on using such methods to extract analytic corrections
to the proximity force approximation:
\begin{itemize}
 \yitem ``Casimir Force for a Sphere in Front of a Plane beyond Proximity Force
  Approximation,'' M.~Bordag and V.~Nikolaev, J. Phys.\ A: Math.\ Gen.\ {\bf41},
164002 (10 pages) (2008). (A)
\yitem ``First Analytic Correction Beyond the Proximity Force Approximation
  in the Casimir Effect for the Electromagnetic Field in Sphere-Plane
  Geometry,'' M.~Bordag and V.~Nikolaev, Phys.\ Rev.\ D  {\bf81}, 065011 (14 pages) (2010). (A)
\end{itemize}
Although numerical and analytic methods agree for Dirichlet boundary conditions,
discrepancies persist for Neumann and electromagnetic boundaries. See Refs.~[\ref{emigjsm}],
[\ref{mnlr}],
and the following.
\begin{itemize}
\yitem ``Numerical Evaluation of the Casimir Interaction between Cylinders,''
F.~C. Lombardo, F.~D. Mazzitelli, and P.~I. Villar,
Phys.\ Rev.\ D {\bf78}, 085009 (11 pages) (2008). (A)
\end{itemize}

While the above references are typically concerned with rather small
corrections that might be seen in the current generation of experiments,
it was immediately recognized that much more general geometries were
amenable to calculations:
\begin{itemize}
\yitem ``Casimir Forces between Arbitrary Compact Objects,''
T. Emig, N. Graham, R. L. Jaffe, and M. Kardar,
Phys.\ Rev.\ Lett.\ {\bf99}, 170403 (4 pages) (2007). (I)
\yitem ``Multiple Scattering Methods in Casimir Calculations,''
K. A Milton and  J. Wagner,
J. Phys.\ A: Math.\ Theor.\ {\bf41}, 155402 (24 pages) (2008). (A)
\yitem ``Weak Coupling Casimir Energies for Finite Plate Configurations,''
J. Wagner, K. A. Milton, and  P. Parashar,
J. Phys.: Conf.\ Ser.\ {\bf161}, 012022 (10 pages) (2009). (I)
\end{itemize}

Very interesting modelling of Casimir-Polder forces are provided by
recent calculations of quantum vacuum forces and torques on a 
small object enclosed in a spherical conducting shell (or exterior to
a conducting shell):
\begin{itemize}
\yitem ``Casimir Interactions of an Object Inside a Spherical Metal Shell,''
S. Zaheer, S. J. Rahi, T. Emig, and R. L. Jaffe,
Phys.\ Rev.\ A {\bf 81}, 0305020(R) (4 pages) (2010). (I)
\yitem ``Casimir Potential of a Compact Object Enclosed by  a Spherical 
Cavity,''
S. Zaheer, S. J. Rahi, T. Emig, and R. L. Jaffe,
Phys.\ Rev.\ A {\bf 82}, 052507 (10 pages) (2010). (I)
\end{itemize}

Simultaneously with the development of these analytical techniques,
numerical methods, based on finite-difference solutions of Maxwell's
equations, have been developed by an independent MIT group:
\begin{itemize}
\yitem ``Efficient Computation of Casimir Interactions between Arbitrary 
3D Objects,'' M. T. Homer Reid, A. W. Rodriguez, J. White, and S. G. Johnson,
 Phys.\ Rev.\ Lett.\ {\bf103}, 040401 (4 pages) (2009). (I)
\yitem ``Casimir Forces in the Time Domain II: Applications,''
A. P. McCauley, A. W. Rodriguez, J. D. Joannopoulos, and S. G. Johnson,
 Phys.\ Rev.\ A {\bf81}, 012119 (10 pages) (2010). (A)
\yitem ``Achieving a Strongly Temperature-Dependent Casimir Effect,''
A. W. Rodriguez, D. Woolf, A. P. McCauley, F. Capasso, J. D. Joannopoulos, 
and S. G. Johnson,
Phys.\ Rev.\ Lett.\ {\bf105}, 060401 (4 pages) (2010). (I)
\end{itemize}

\section{Repulsive Casimir forces}
\label{secrep}
Motivated by possible nanotechnological applications, there has been a
resurgence of interest in repulsive Casimir and Casimir-Polder forces.
Of course, it was known by Lifshitz and collaborators that repulsion
could be achieved if two different dielectric materials, of permittivities
$\varepsilon_1$ and $\varepsilon_2$ were separated by a medium with an
intermediate value of the dielectric constant $\varepsilon_1<\varepsilon_3
<\varepsilon_2$.  This has recently be confirmed experimentally:
\begin{itemize}
\yitem ``Measured Long-Range Repulsive Casimir-Lifshitz Forces,''
J. N. Munday, F.  Capasso, 
and V. A.  Parsegian, Nature {\bf 457}, 170--173 (2009).
 (I) 
\end{itemize}
Precursors of this experiment should also be mentioned.
\begin{itemize}
 \yitem ``Direct Measurement of Repulsive van der Waals Interactions Using an Atomic Force Microscope,''
A.~Milling, P.~Mulvaney, and I.~Larson, J.\ Colloid Interf.\ Sci.\ {\bf 180}, 460--465 (1996). (I)
\yitem  ``Direct Measurement of Repulsive and Attractive van der Waals Forces between Inorganic Materials,''
A.~Meurk, P.~F.\ Luckham, and L.~Bergstrom, Langmuir {\bf 13}, 3896--3899 (1997). (I)
\yitem ``Repulsive van der Waals Forces for Silica and Alumina,''
 S.\ Lee and W.~M.\ Sigmund, J.\ Colloid Interf.\ Sci.\ {\bf 243}, 365--369 (2001). (I)
\yitem ``AFM Study of Repulsive van der Waals Forces between Teflon AF$^{\rm TM}$ 
Thin Film and Silica or Alumina,'' 
S.\ Lee and W.~M.\ Sigmund, Colloids Surf.\ A {\bf 204}, 43--50 (2002). (I)
\yitem ``Superlubricity Using Repulsive van der Waals Forces,''
A.~A.\ Feiler, L.\ Bergstr\"{o}m, and M.~W.\ Rutland, Langmuir {\bf 24}, 2274--2276 (2008). (I)
\end{itemize}

More interestingly, Boyer demonstrated in 1968 that the Casimir self-stress
on a sphere was repulsive,  whereas that for a conducting circular cylindrical
shell is attractive.
\begin{itemize}
\yitem \label{boyer68}
``Quantum Electromagnetic Zero-Point Energy of a Conducting Spherical 
Shell and the Casimir Model for a Charged Particle,''
T. H. Boyer, Phys. Rev. {\bf174}, 1764--1776 (1968). (I)
\yitem \label{balian78}
``Electromagnetic Waves Near Perfect Conductors. II. Casimir Effect,''
R. Balian and B. Duplantier, Ann.\ Phys.\ (N.Y.) {\bf 112}, 165--208 (1978). 
(I)
\yitem \label{milton78}
``Casimir Self-Stress on a Perfectly Conducting Spherical Shell,''
K. A. Milton, L. L. DeRaad, Jr., and J. Schwinger, Ann.\ Phys.\ (N.Y.)
{\bf 115}, 388--403 (1978). (I)
\yitem \label{deraad81}
``Casimir Self-Stress on a Perfectly Conducting Cylindrical
Shell,'' L. L. DeRaad, Jr.\ and K. A. Milton, Ann.\ Phys.\ (N.Y.) {\bf136},
229--242. (A)
\end{itemize}

Casimir-Polder forces could be repulsive if they involved magnetic moment
couplings, rather than the usual electric-dipole couplings:
\begin{itemize}
\yitem ``Theory of Casimir-Polder Forces,''
B.-S. Skagerstam, P. K. Rekdal,  and A. H. Vaskinn,
Phys.\ Rev.\ A {\bf 80}, 022902 (12 pages) (2009). (A)
\end{itemize}  This is consistent with the result (\ref{gencp}).
In fact, many years ago Boyer showed that Casimir force between
a perfect electrically conducting plate and perfect magnetically
conducting plate (that is, the permittivity $\varepsilon\to\infty$,
and the permeability $\mu\to\infty$, respectively) is repulsive.
\begin{itemize}
 \yitem  ``Van der Waals Forces and Zero-Point Energy for Dielectric 
and Permeable Materials.'' T. H. Boyer,
Phys.\ Rev.\ A {\bf9}, 2078--2084 (1974). (I)
\end{itemize}
This suggests that properly designed metamaterials might exhibit 
repulsive forces, but the severe difficulty is that the unusual
magnetic properties must persist over a very wide frequency range
to achieve an observable effect, since the reflection coefficients
appear inside an integral.
\begin{itemize}
\yitem ``Casimir Force between Designed Materials: What Is Possible and What Not,'' 
C. Henkel and K. Joulain, Europhys.\ Lett. {\bf72}, 929--935 (2005). (I)
\yitem ``Casimir Repulsion and Metamaterials,''
I. G. Pirozhenko and A. Lambrecht, J. Phys. A 41, 164015 (8 pages) (2008). (A)
\yitem ``Casimir-Lifshitz Theory and Metamaterials,''
F. S. S. Rosa, D. A. R. Dalvit, and P. W. Milonni, Phys.\ Rev.\ Lett.\ {\bf100}, 183602 (4 pages) (2008). (I)
\yitem ``First-Principles Study of Casimir Repulsion in Metamaterials,''
V. Yannopapas and N. V. Vitanov, Phys.\ Rev.\ Lett.\ {\bf103}, 120401 (4 pages) (2009). (I)
\yitem `` Physical Restrictions on the Casimir Interaction of Metal-Dielectric Metamaterials: 
An Effective-Medium Approach,''
M. G. Silveirinha and S. I. Maslovski, Phys.\ Rev.\ A {\bf82}, 052508 (5 pages) (2010). (A)
\yitem ``Microstructure Effects for Casimir Forces in Chiral Metamaterials,''
McCauley et al., Phys.\ Rev.\ B {\bf82}, 165108 (5 pages) (2010). (I)
\end{itemize}

It also might be possible to achieve repulsion by focusing electromagnetic
fields with parabolic mirrors:
\begin{itemize}
\yitem ``Focusing Vacuum Fluctuations. II,'' L. H. Ford and N. F. Svaiter,
Phys.\ Rev.\ A {\bf 66}, 062106 (13 pages) (2002). (I)
\end{itemize}

Very recently, numerical calculations have shown that an elongated
cylinder directly above a circular hole in a conducting sheet experiences
a Casimir force that is attractive at large distances but is repulsive
at distances that are of the order of the diameter of the hole.  The
equilibrium point is, of course, unstable, since the energy is lowered by
moving the object slightly off-axis.
\begin{itemize}
\yitem ``Casimir Repulsion between Metallic Objects in Vacuum,''
M. Levin, A. P. McCauley, A. W. Rodriguez, M. T. Homer Reid, and S. G. Johnson,
Phys.\ Rev.\ Lett.\ {\bf105}, 090403 (4 pages) (2010). (A)
\yitem ``A Diagrammatic Expansion of the Casimir Energy in Multiple 
Reflections: Theory and Applications,'' M. F. Maghrebi,
Phys.\ Rev.\ D {\bf83}, 045004 (12 pages) (2011). (I)
\end{itemize}
The latter reference verifies the repulsion by analytical calculation.
Presumably, the same repulsion occurs for the Casimir-Polder force for
an anisotropic molecule above a punctured conducting plate; calculations 
are currently underway to investigate this effect.

\section{Self-energies}\label{secse}
We have already referred to several instances of Casimir self-energies,
for example, for spheres and cylinders, Refs.~[\ref{boyer68}], [\ref{balian78}]
[\ref{milton78}], [\ref{deraad81}]. See also
\begin{itemize}
 \yitem ``Casimir Energies for Massive Scalar Fields in a Spherical Geometry,''
 M. Bordag, E. Elizalde, K. Kirsten, and S. Leseduarte, 
Phys.\ Rev.\ {\bf D56}, 4896--4904 (1997). (A)
\yitem  Casimir Energy for a Massive Fermionic Quantum Field with a
Spherical Boundary,
E. Elizalde, M. Bordag, and K. Kirsten, J. Phys.\ A {\bf 31}, 1743--1759 (1998). (A)
\end{itemize}
There was also extremely interesting results on the dimensional
dependence of the Casimir effect, for example,
\begin{itemize}
 \yitem ``Scalar Casimir Effect for a $D$-Dimensional Sphere,''
C. M. Bender and K. A. Milton,
Phys.\ Rev.\ D {\bf50}, 6547--6555 (1994). (A)
\yitem ``Vector Casimir Effect for a $D$-Dimensional Sphere,''
K. A. Milton, Phys.\ Rev.\ D {\bf 55}, 4940--4946 (1997). (A)
\end{itemize}

 Such calculations have somewhat obscure
physical meaning, for although they are mathematically well-defined, it is
difficult to see how they lead to an observable effect.  For example, if
a conducting sphere is cut in half and pulled apart, it will experience an
attraction (due to the closest parts of the surface) not a repulsion.
Nevertheless, the study of self-energies has remained an important part of
Casimir studies.  For a status report on the subject see  Ref.~[\ref{milton11}].
Very recent work that shows intriguing systematics of the sign and magnitude
of the Casimir self-stress is given in the following.
\begin{itemize}
 \yitem ``Casimir Energies of Cylinders: Universal Function,''
E. K. Abalo, K. A. Milton, and L. Kaplan,
Phys.\ Rev.\ D {\bf82}, 125007 (12 pages) (2010). (I)
\end{itemize}

\section{New experimental developments}
Within the last year, the first experiment was published measuring
the Casimir-Polder force between an atom and a solid surface in the
intermediate region between the nonretarded (vdW) and the retarded (CP) 
regime.
\begin{itemize}
\yitem ``Direct Measurement of Intermediate-Range Casimir-Polder Potentials,''
H. Bender, Ph. W. Courteille, C. Marzok, C. Zimmermann, and S. Slama,
Phys.\ Rev.\ Lett.\ {\bf 104}, 083201 (4 pages) (2010). (I)
\end{itemize}
The following paper shows that the data agree best with a full QED calculation,
and not with the retarded CP potential.
\begin{itemize}
\yitem ``Macroscopic QED---Concepts and Applications,''
S. Scheel and S. Y. Buhmann, Acta Phys.\ Slov.\ {\bf58}, 675--809
(2008). (I)
\end{itemize}

There are new experiments that will test various aspects of Casimir-Polder
forces, for example, between an atom and a complex surface.
\begin{itemize}
\yitem ``Nanowire Atomchip Traps for Sub-Micron Atom-Surface Distances,'' 
R. Salem, Y. Japha, J. Chab\'e, B. Hadad, M. Keil, K. A. Milton, 
and R. Folman, 
New J. Phys.\ {\bf12}, 023039 (28 pp) (2010). (I)
\end{itemize}
They discuss a proposed  measurement of atoms trapped in a Bose-Einstein 
condensate above a trapping wire on bilayer dielectric surface.
And the CP interaction of a atom with a wall could be tested with atomic
clocks:
\begin{itemize}
\yitem ``Mapping Out Atom-Wall Interaction with Atomic Clocks,''
A. Derevianko, B. Obreshkov, and V. A. Dzuba, Phys.\ Rev.\ Lett.\ {\bf 103},
133201 (4 pages) (2009). (I)
\end{itemize}
Further work will allow us to calculate energy shifts of an atom near a
layered microstructure:
\begin{itemize}
\yitem ``Casimir-Polder Interaction between an Atom and a Dielectric Slab,''
A. M. Contreras Reyes and C. Eberlein, Phys.\ Rev.\ A {\bf 80}, 032901
(10 pages) (2009). (I)
\end{itemize}

\section{Conclusions}
\label{concl}
This Resource Letter is designed to guide the reader, whether a student or
a researcher in another field who desires to become initiated in the subject,
to the extensive 
literature on Casimir-Polder, van der Waals, and in general quantum
vacuum forces between neutral atoms, molecules, and microscopic and macroscopic
bodies. (For example, Physical Review lists nearly 800 citations of
the 1948 Casimir-Polder paper, Ref.~[\ref{cppaper}].)
 As one can tell, the field is undergoing a rapid period of 
development, with important innovations in both theory and experiment.
It is nearly certain that within the next decade Casimir forces will give
rise to practical nanotechnological applications.

But there is more.  The current cosmological observations strongly suggest
that 70\% of the energy in the universe consists of ``dark energy,'' which
results in the observed cosmic acceleration.  Although the errors are large,
the equation of state of the dark energy, $w=-p/\rho$, in terms of the pressure
and density of the dark energy, is consistent with the value predicted if
the dark energy were Einstein's cosmological constant, $w=-1$.  If so, it
is highly likely that the dark energy is due to quantum fluctuations of
fields in the universe or extra dimensions.

For a Resource Letter on Dark Energy see
\begin{itemize}
\yitem ``Resource Letter DEAU-1: Dark Energy and the Accelerating Universe,''
E. V. Linder, Am.\ J. Phys.\ {\bf76}, 197--204 (2008). (E)
\end{itemize}
For ideas of how dark energy might arise from quantum fluctuations, see
for example,
\begin{itemize}
\yitem  ``Constraints on Extra Dimensions from Cosmological and Terrestrial 
Measurements,'' K. A. Milton, Grav.\ Cosmol.\ {\bf8}, 65--72 (2002). (I) 
\end{itemize}
However, the particular model proposed there seems incompatible with the latest
laboratory tests on gravity at short distance.
For a survey of the long-standing difficulties in understanding the 
cosmological constant, see
\begin{itemize}
\yitem ``The Cosmological Constant Problem,'' S. Weinberg, 
Rev.\ Mod.\ Phys.\ {\bf61}, 1--23 (1989). (I)
\end{itemize}
This is updated in
\begin{itemize}
\yitem ``The Cosmological Constant Problems,''
S. Weinberg, talk given at the
4th International Symposium on Sources and Detection of Dark Matter in the 
Universe (DM 2000) 23--25 February 2000, Marina del Rey, California,
Proceedings, ed.\ D. B. Cline. (Berlin,  Springer-Verlag, 2001);
arXiv:astro-ph/0002387. (I)
\end{itemize}

\acknowledgments
The research underlying the work by the author reported here was supported
in part by grants from the US National Science Foundation and the US
Department of Energy. I want to thank many colleagues for their very useful
suggestions of references.
\end{document}